\documentclass[10pt,usletter]{article}
  \usepackage[totalwidth=420pt,totalheight=625pt]{geometry}

\newcommand{\cv}[1]{}  \newcommand{\av}[1]{#1} 

\newcommand{\SPACE}{\cv{\vspace{-0.4mm}}}

\usepackage[T1]{fontenc}
\usepackage{graphicx}
\usepackage{amsmath,amssymb}
\usepackage{xspace}
\usepackage{booktabs}
\usepackage{array}
\usepackage{tikz}
\usetikzlibrary{positioning,shapes,arrows.meta}

\usepackage{xcolor}
  \usepackage{url,hyperref}

\usepackage{listings}
\lstdefinelanguage{lean}{
  morekeywords={theorem,def,lemma,example,where,let,in,if,then,else,match,with,do,return,have,show,by,sorry,import,open,namespace,end,section,variable,instance,class,structure,inductive,noncomputable,deriving,extends,abbrev,axiom,private,protected,mutual,attribute,macro,syntax,notation,prefix,infix,postfix,scoped,local,calc,fun,forall,exists,Type,Prop,Sort,true,false},
  sensitive=true,
  morecomment=[l]{--},
  morecomment=[s]{/-}{-/},
  morestring=[b]",
}
\newcommand{\veripb}{\textsc{VeriPB}\xspace}
\newcommand{\cakepb}{\textsc{CakePB}\xspace}
\newcommand{\roundingsat}{\textsc{RoundingSat}\xspace}

\newcommand{\lrat}{\textsc{LRAT}\xspace}

\newcommand{\pblean}{\textsc{PBLean}\xspace}
\newcommand{\aprx}{{\raise.17ex\hbox{$\scriptstyle\sim$}}}

\begin{document}
\cv{\title{\pblean: Pseudo-Boolean Proof Certificates for Lean~4}}
\av{\title{\pblean: Pseudo-Boolean Proof Certificates for Lean~4}}
\author{Stefan Szeider\\[4pt]
    \small Algorithms and Complexity Group\\[-3pt]
    \small TU Wien, Vienna, Austria\\[-3pt]
    \small \href{https://www.ac.tuwien.ac.at/people/szeider/}{www.ac.tuwien.ac.at/people/szeider/}
  }
  \date{}
\maketitle
\begin{abstract} \sloppypar We present \pblean, a method for importing \veripb pseudo-Boolean
  (PB) proof certificates into Lean~4. Key to our approach is
  \emph{reflection}: a Boolean checker function whose soundness is
  fully proved in Lean and executed as compiled native code. Our
  method scales to proofs with tens of thousands of steps that would
  exhaust memory under explicit proof-term construction. Our checker
  supports all \veripb kernel rules, including cutting-plane
  derivations, proof-by-contradiction subproofs, and redundance-based
  reasoning for symmetry breaking. In contrast to
  external verified checkers that produce verdicts, our integration
  yields Lean theorems that can serve as composable lemmas in larger
  formal developments. To derive theorems about the original
  combinatorial problems rather than about PB constraints alone, we
  support verified encodings. This closes the trust gap between solver
  output and problem semantics since the constraint translation and
  its correctness proof are both formalized in Lean.  We demonstrate
  the approach on various combinatorial problems.

\end{abstract}

\section{Introduction} \av{\thispagestyle{empty}}
\SPACE
\label{sec:intro}

Pseudo-Boolean (PB) solvers handle linear constraints over Boolean
variables and have become prominent tools for combinatorial optimization
and verification~\cite{ElffersN18}.
\av{Since these solvers carry out
sophisticated derivations, independent verification of their results is
particularly desirable. However, PB reasoning involves}\cv{Independently verifying their results is therefore important,
but PB reasoning involves}
cutting-planes proof steps that resolution-based formats cannot capture
efficiently.

\sloppypar Haken's classical result~\cite{Haken85} shows that the
pigeonhole principle requires exponential-size resolution proofs, whereas
polynomial-size cutting-plane proofs exist~\cite{CookCT87}. Beyond
pigeonhole, cutting planes also efficiently capture XOR/parity
constraints~\cite{GochtN21} and certified symmetry
breaking~\cite{BogaertsGMN22}, which are reasoning methods that lack
efficient resolution-based proof logging. This motivated the \veripb
proof format~\cite{ElffersGMN20,BogaertsGMN22}, based on cutting
planes and now used by solvers such as \roundingsat for certified PB
solving.

The \lrat format~\cite{HeuleHKW17} enables SAT solvers to produce
clausal certificates that can be checked in theorem provers. Lean's
\texttt{FromLRAT} module~\cite{FromLRAT} imports such certificates and
produces kernel-checked theorems. Heule and
Scheucher~\cite{HeuleS24} established the empty hexagon theorem using
SAT solving with verified certificate checking; Subercaseaux
et~al.~\cite{SubercaseauxNGCCH24} subsequently formalized this
encoding in Lean and verified the reduction from geometry to SAT. A
similar integration for pseudo-Boolean reasoning has
been missing.

\av{Why not compile PB reasoning to clausal proofs and reuse
\lrat? Bryant et al.~\cite{BryantBH22} showed that this is possible
via BDD compilation, and their approach handles classical hard cases
such as pigeonhole and parity efficiently. However, the compilation
requires coordinating a PB solver with a BDD-based clausal proof
generator, and whether it preserves efficiency for all PB reasoning
remains open. Direct \veripb import avoids this detour.}\cv{One might compile PB reasoning to clausal proofs and reuse
\lrat~\cite{BryantBH22}, but this requires coordinating a PB solver
with a BDD-based clausal proof generator, and whether it preserves
efficiency for all PB reasoning remains open. Direct \veripb import
avoids this detour.}

\av{The \cakepb project~\cite{GochtMMNOT24} provides a verified checker
extracted from HOL4 via CakeML, producing a standalone binary that
outputs verdicts. The main advantage of \cakepb is its small, trusted
computing base enabled by verified compilation. However, the verified
result (a binary verdict) cannot serve as a
lemma in larger formal developments.}\cv{The \cakepb project~\cite{GochtMMNOT24} provides a verified checker
extracted from HOL4 via CakeML, producing a standalone binary. While
this yields a small trusted computing base, the verified result (a
binary verdict) cannot serve as a lemma in larger formal developments.}

A proof certificate establishes that certain PB constraints are
unsatisfiable, but says nothing about whether those constraints correctly
model the intended problem. An encoding bug can produce unsatisfiable
constraints even when solutions exist. We address this gap with
\emph{verified encodings}: both the constraint translation and its
correctness proof are entirely in Lean, so that certificate import yields
theorems about the original problem, not merely about PB constraints.

We present \pblean, a \veripb integration for Lean~4 that parses kernel-format
proofs and verifies them via \emph{reflection}. Instead of building
Lean proof terms for each derivation step, we define a Boolean checker
function and prove its soundness once; Lean's \texttt{native\_decide}
then runs the checker as compiled code. This follows the pattern of
Lean's \texttt{bv\_decide} tactic~\cite{LeanBVDecide} and scales to
instances that would exhaust memory under explicit proof construction.
Our implementation covers all \veripb kernel rules, including
proof-by-contradiction subproofs for optimization reasoning and
redundance/dominance rules for symmetry breaking. We apply
the verified encoding workflow to independence numbers of Paley
graphs and other combinatorial theorems to obtain end-to-end-verified
results.

\section{Background} \label{sec:background}
\SPACE
A pseudo-Boolean (PB) constraint has the form
$\sum_{i=1}^n a_i \ell_i \geq d$,
where $a_i, d \in \mathbb{N}$ are non-negative coefficients, and each
$\ell_i$ is a literal (a variable $x_i$ or its negation $\overline{x}_i$).
Under an assignment $\sigma$, a literal evaluates to~1 if satisfied
and~0 otherwise. The constraint is satisfied when the weighted sum
reaches the degree~$d$. For example, $3x_1 + 5x_2 \geq 4$ is
satisfied by $\sigma(x_1) = \sigma(x_2) = 1$ (sum $= 8 \geq 4$) but
not by $\sigma(x_1) = 1$, $\sigma(x_2) = 0$ (sum $= 3 < 4$). Clauses
are a special case: $x_1 \vee x_2 \vee
\overline{x}_3$ corresponds to $x_1 + x_2 + \overline{x}_3 \geq 1$.

Cutting-planes reasoning derives new constraints via:
(1)~addition of constraints,
(2)~scalar multiplication by $k \in \mathbb{N}^+$,
(3)~division by $k$ with the degree rounded up, and
(4)~saturation, which caps each coefficient at the degree.
These rules strictly generalize resolution~\cite{CookCT87}: resolution
corresponds to addition with complementary literal cancellation. The
power of cutting planes stems from the
interplay of multiplication and division, which derives constraints
unreachable by resolution alone~\cite{VinyalsEGGN18}.

The \veripb proof format~\cite{ElffersGMN20,BogaertsGMN22} provides
two modes: an augmented mode with high-level rules (symmetry breaking,
redundance-based strengthening) and a kernel mode (version 3.0) with only primitive operations. The kernel format's small rule set of just eight rule types enables a rigorous soundness proof and makes it well-suited for theorem prover integration. It supports:
\texttt{pol} (polynomial operations including all cutting-planes rules),
\texttt{rup} (reverse unit propagation with conflict hints),
\texttt{pbc} (proof by contradiction with explicit subproofs),
\texttt{red}/\texttt{dom} (redundance and dominance, adding equisatisfiable constraints via a substitution witness),
\texttt{del} (constraint deletion, sound by monotonicity),
\texttt{weaken} (coefficient weakening),
\texttt{sol}/\texttt{soli} (solution logging), and
\texttt{conclusion} (UNSAT/SAT/BOUNDS verdict).
Solvers such as \roundingsat produce augmented proofs;
\texttt{veripb --elaborate} translates them to kernel format.

Lean~4~\cite{MouraU21} provides a metaprogramming framework where
tactics run in monads such as \texttt{MetaM}. Code in \texttt{MetaM}
constructs \texttt{Expr} values (Lean's internal representation of
proof terms), and, upon type-checking, Lean's kernel verifies their correctness.
Our approach builds \texttt{Expr} values from \veripb proofs; the
parser and proof-term constructor are untrusted code, as in
\texttt{FromLRAT}. The trusted base consists of Lean's kernel and the
soundness lemmas presented in Section~\ref{sec:formalization}.

\section{Lean Formalization} \label{sec:formalization}
\SPACE
For constraint representation, we use natural number arithmetic
throughout. A $\mathit{Literal}$ is $\mathtt{pos}\ i$ or
$\mathtt{neg}\ i$, and a $\mathit{Term}$ is a pair
$(a, \ell) \in \mathbb{N} \times \mathit{Literal}$; a $\mathit{Constr}$
is a term list plus degree. Given a valuation
$v : \mathbb{N} \to \mathit{Bool}$, we define
$\mathit{evalLit}(v, \ell) \in \{0,1\}$ (1 if satisfied) and
$\mathit{evalSum}(v, T) = \sum_{(a,\ell) \in T} a \cdot \mathit{evalLit}(v,\ell)$
recursively to support structural induction;
$\mathit{coeffSum}(T) = \sum_{(a,\ell) \in T} a$ is the sum of coefficients. A constraint $(T, d)$ is
satisfied when $d \leq \mathit{evalSum}(v, T)$.

Our 15 soundness lemmas are summarized in Table~\ref{tab:lemmas}; every
lemma is proved in Lean and verified by the kernel. Two details matter
for matching \veripb's semantics: normalization must precede saturation
and division, and the negation of $\sum a_i \ell_i \geq d$ is
$\sum a_i \bar\ell_i \geq (\sum a_i) - d + 1$.

\begin{table}[tbh]
\centering
\caption{Soundness lemmas for \veripb kernel operations.}
\label{tab:lemmas}
\av{\medskip}
\cv{\small}
\begin{tabular}{@{}l>{\raggedright\arraybackslash}p{9.0cm}@{}}
\toprule
Operation & Lean Type \\
\midrule
\multicolumn{2}{@{}l}{\textit{Cutting planes}} \\
\texttt{add\_sat} &
  $c_1.\mathit{sat}\ v \to c_2.\mathit{sat}\ v \to (c_1.\mathit{terms} \mathbin{++} c_2.\mathit{terms},\ c_1.\mathit{deg} + c_2.\mathit{deg}).\mathit{sat}\ v$ \\
\texttt{mul\_sat} &
  $c.\mathit{sat}\ v \to (\mathit{map}\ (\lambda (a,\ell).\ (k \cdot a, \ell))\ c.\mathit{terms},\ k \cdot c.\mathit{deg}).\mathit{sat}\ v$ \\
\texttt{div\_sat} &
  $0 < k \to c.\mathit{sat}\ v \to (\mathit{map}\ (\lambda (a,\ell).\ (\lceil a/k \rceil, \ell))\ c.\mathit{terms},\ \lceil c.\mathit{deg}/k \rceil).\mathit{sat}\ v$ \\
\texttt{saturate\_sat} &
  $c.\mathit{sat}\ v \to (\mathit{map}\ (\lambda (a,\ell).\ (\min(a, c.\mathit{deg}), \ell))\ c.\mathit{terms},\ c.\mathit{deg}).\mathit{sat}\ v$ \\
\midrule
\multicolumn{2}{@{}l}{\textit{Normalization}} \\
\texttt{cancel\_pair\_sat} &
  $\min(a,b) \leq d \to (\ldots (a,\ell) \ldots (b,\bar\ell) \ldots,\ d).\mathit{sat}\ v
  \to (\ldots (a{-}m,\ell) \ldots (b{-}m,\bar\ell) \ldots,\ d{-}m).\mathit{sat}\ v$ \\
\texttt{remove\_zero\_sat} &
  $(\mathit{pre} \mathbin{++} (0,\ell){::}\mathit{post},\ d).\mathit{sat}\ v \to (\mathit{pre} \mathbin{++} \mathit{post},\ d).\mathit{sat}\ v$ \\
\texttt{merge\_terms\_sat} &
  $(\ldots (a,\ell) \ldots (b,\ell) \ldots,\ d).\mathit{sat}\ v
  \to (\ldots (a{+}b,\ell) \ldots,\ d).\mathit{sat}\ v$ \\
\midrule
\multicolumn{2}{@{}l}{\textit{Axioms and contradiction}} \\
\texttt{lit\_axiom\_pos/neg} &
  $\forall v.\ ([(1, \ell)],\ 0).\mathit{sat}\ v$ \\
\texttt{contra\_unsat} &
  $\mathit{coeffSum}(c) < c.\mathit{deg} \to \neg\ c.\mathit{sat}\ v$ \\
\midrule
\multicolumn{2}{@{}l}{\textit{Negation and proof by contradiction}} \\
\texttt{negate\_sat\_of\_not} &
  $c.\mathit{deg} \leq \mathit{coeffSum}(c) \to \neg\ c.\mathit{sat}\ v \to c.\mathit{negate}.\mathit{sat}\ v$ \\
\texttt{propagation\_forces} &
  $(T,d).\mathit{sat}\ v \to \mathit{coeffSum}(T \setminus \{(a,\ell)\}) < d \to \mathit{evalLit}(v,\ell) = 1$ \\
\texttt{pbc\_sound} &
  $(\forall v.\ F.\mathit{allSat}\ v \to C.\mathit{negate}.\mathit{sat}\ v \to \bot)
  \to F.\mathit{proof}\ C$ \\
\midrule
\multicolumn{2}{@{}l}{\textit{Redundance (red/dom)}} \\
\texttt{applySubstConstr\_sat\_rev} &
  $C|_\omega.\mathit{sat}\ v \to C.\mathit{sat}\ (\omega(v))$ \\
\texttt{constr\_sat\_noSubst} &
  $\mathit{unaffected}(\omega, C) \to C.\mathit{sat}\ v \to C.\mathit{sat}\ (\omega(v))$ \\
\bottomrule
\end{tabular}
\end{table}

We give two verification approaches with complementary trade-offs.

\paragraph{Explicit proof terms.}
The first approach builds Lean \texttt{Expr} terms by composing
soundness lemmas. For \texttt{pol} steps, each token in the
reverse Polish notation (RPN) expression produces an application: ID lookups retrieve earlier proofs,
addition applies \texttt{add\_sat}, multiplication applies
\texttt{mul\_sat}, and so on. For \texttt{rup} steps, we negate the
target, run PB unit propagation to find a conflicting hint, and then
build a proof accumulator. For \texttt{pbc} subproofs, we extend the
context with the negated assumption and verify inner steps recursively.
The resulting \texttt{Expr} is then type-checked by Lean's kernel,
producing a theorem. Its trusted base consists of just the kernel and
the soundness lemmas. This approach has the disadvantage that the
accumulated proof terms exhaust memory for instances with thousands of
constraints.

\paragraph{Reflection.}
To overcome this limitation, our second approach follows the pattern of
Lean's \texttt{bv\_decide}~\cite{LeanBVDecide}. We define a Boolean
function $\mathit{check}(F, \pi)$ that takes a constraint set~$F$ and
a proof string~$\pi$, replays all derivation steps, and returns
$\mathit{true}$ iff a contradictory constraint is derived. We then
prove a soundness theorem: if $\mathit{check}(F, \pi) = \mathit{true}$
then $F$ is unsatisfiable. This theorem is fully proved in Lean
without custom axioms. Lean's \texttt{native\_decide} runs the checker
as compiled code, yielding a kernel proof term consisting solely of
the soundness theorem applied to the computation witness. The resulting theorem
has the same trust basis as \texttt{bv\_decide} and
\texttt{omega}---namely, Lean's standard axioms (\texttt{propext},
\texttt{Classical.choice}, \texttt{Quot.sound}) plus
\texttt{Lean.trustCompiler}, and scales to proofs with thousands of
steps.

\paragraph{Trust model.}
The reflection-based checker relies on
\texttt{Lean.trustCompiler}, the axiom that Lean's compiled native
code agrees with its kernel evaluator. This is the same trust
assumption underlying \texttt{bv\_decide}, \texttt{native\_decide},
and \texttt{omega}. Lean's kernel does not verify native compilation;
a bug in the compiler could in principle produce an incorrect
evaluation result.  In practice, the Lean compiler is mature and
widely used, and the alternative---explicit proof-term
construction---does not scale beyond small instances.
The explicit proof-term approach (our first method above) avoids this
axiom entirely, providing a fallback for instances where
\texttt{Lean.trustCompiler} is unacceptable.

\section{Implementation} \label{sec:implementation}
\SPACE
\pblean is a standalone Lean~4 project (v4.28.0-rc1) with no
Mathlib dependency. The kernel layer
(\texttt{PseudoBoolean.lean}, \aprx900 lines) consists of types and 15
soundness lemmas. The code is available on
GitHub.\footnote{\url{https://github.com/leansolving/pblean}}

The reflection-based checker (\texttt{Reflect.lean}, \aprx2500 lines)
uses a line-by-line tokenizer to convert \veripb proofs into an
abstract syntax tree, handling nested \texttt{pbc} subproofs
recursively. It maintains a constraint database organized as a hash
map, with an \texttt{@[implemented\_by]} annotation providing an
array-based runtime replacement for performance without weakening the
proof. Each rule updates or queries this database: \texttt{pol}
evaluates the RPN expression, \texttt{rup} derives a conflict by
negating the target and applying PB unit propagation, \texttt{pbc}
saves the database, adds the negated assumption, verifies inner steps
recursively, then restores the snapshot, and \texttt{red}/\texttt{dom}
verify coverage of affected constraints and process subproofs for each
proof goal before adding the equisatisfiable constraint. The checker returns
\texttt{true} only when a contradictory constraint arises, and we
establish soundness by induction over proof steps, connecting the
Boolean computation to the kernel lemmas.

To exploit the integration, one provides an OPB file and a
kernel-format proof file. The command \texttt{veripb\_reflect} parses
both, runs the checker via \texttt{native\_decide}, and registers the
resulting unsatisfiability theorem under a given name. For
domain-specific applications, wrapper commands such as
\texttt{independent\_set\_reflect} additionally generate the encoding
from a problem instance and compose the bridge theorem automatically.

\begin{figure}[tbh]
\centering
\begin{tikzpicture}[
    node distance=0.3cm and 0.45cm,
    box/.style={draw, rounded corners, minimum height=1.0cm, minimum width=1.8cm, align=center, font=\small},
    trusted/.style={box, fill=blue!10},
    untrusted/.style={box, fill=gray!15, dashed},
    artifact/.style={draw, rounded corners=2pt, minimum height=0.6cm, minimum width=1.2cm, align=center, font=\footnotesize, fill=yellow!20},
    arr/.style={-Stealth, thick},
    label/.style={font=\scriptsize, text=black!70}
]

\node[trusted] (claim) {Problem\\Instance};

\node[trusted, right=of claim] (encode) {Trusted\\Encoding};

\node[untrusted, right=of encode] (solver) {PB Solver\\+ VeriPB};

\node[trusted, right=of solver] (checker) {Reflection\\Checker};

\node[trusted, right=of checker] (theorem) {Lean\\Theorem};

\node[artifact, below=0.5cm of encode] (opb) {OPB};
\node[artifact, below=0.5cm of solver] (proof) {Kernel Proof};

\draw[arr] (claim) -- (encode);
\draw[arr] (encode) -- (opb);
\draw[arr] (opb) -- (solver);
\draw[arr] (solver) -- (proof);
\draw[arr] (proof) -- (checker);
\draw[arr] (checker) -- (theorem);

\draw[arr, densely dotted, bend left=28] (encode) to node[above, label] {bridge} (theorem);

\node[label, above=0.05cm of solver] {\textit{untrusted}};

\end{tikzpicture}
\caption{End-to-end workflow for trusted encodings. OPB is the standard
  input format for PB~solvers. Solid
  boxes are verified in Lean; the dashed box is untrusted. The dotted
  arrow represents the bridge theorem composing encoding correctness
  with the PB refutation.}
\label{fig:workflow}
\end{figure}
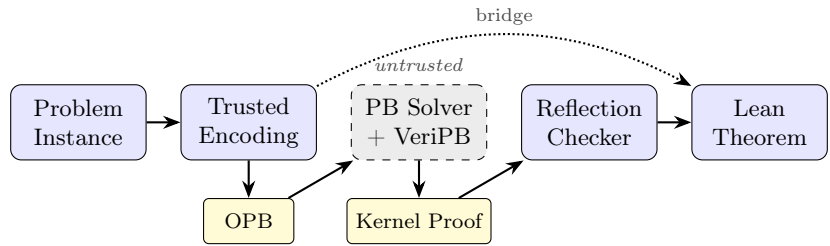

\SPACE
We demonstrate our approach on a problem where the encoding is also
formalized in Lean, closing a trust gap. For SAT, encoding
correctness was only recently formalized~\cite{SubercaseauxNGCCH24};
for PB reasoning, no such formalization existed.
Figure~\ref{fig:workflow} shows the overall workflow.

A verified encoding consists of three components: (1)~an \texttt{encode}
function translating problem instances to PB constraints, (2)~a soundness
theorem proving that unsatisfiability of the encoding implies non-existence
of solutions, and (3)~a bridge theorem composing encoding correctness with
the imported \veripb certificate to yield a theorem about the original problem.

As an example, we consider the independence number (the size of a largest
independent set) of Paley graphs. For a prime $p \equiv 1 \pmod{4}$,
the Paley graph $\mathrm{Paley}(p)$ is defined as the graph with the
vertex set $\mathbb{Z}_p$ and edges $(i,j)$ whenever $j-i$ is a nonzero quadratic
residue. These graphs are self-complementary with an independence number
$\alpha$ of at most $\sqrt{p}$ by the Hoffman eigenvalue bound.  The
encoding for ``does $G$ have an independent set of size $\geq k$?''
uses
\av{\begin{align*}
\forall (u,v) \in E(G) &: \bar{x}_u + \bar{x}_v \geq 1 \quad\text{(independence)}\\
&: x_1 + \cdots + x_n \geq k \quad\text{(cardinality)}
    \end{align*}}
  \cv{$\forall (u,v) \in E(G)\colon \bar{x}_u + \bar{x}_v \geq 1$
    (independence) and $x_1 + \cdots + x_n \geq k$ (cardinality).}
  
The Lean formalization establishes that if these constraints are
unsatisfiable, then $G$ has no independent set of size $\geq k$. Combining
the \veripb certificate with this result yields a theorem about the
graph. The \texttt{independent\_set\_decide} command automates the whole
pipeline: given a graph and a target size~$k$, it produces a Lean
theorem of the form $\lnot\mathit{hasIndependentSet}(G, k)$.

\section{Evaluation} \label{sec:evaluation}
\SPACE
We evaluate \pblean by measuring wall-clock time on an
Apple M2 with cached \texttt{.olean} files. Measured times include parsing, checker
execution via \texttt{native\_decide}, and kernel verification. We use
\roundingsat for solving and \texttt{veripb --elaborate} to produce
kernel-format proofs.

Table~\ref{tab:benchmarks} shows results for independence numbers of
$\mathrm{Paley}(p)$ for primes $p \equiv 1 \pmod{4}$ from 13 to 101.\cv{\footnote{A noteworthy byproduct of this evaluation is a
formally verified proof that $\alpha(\mathrm{Paley}(p))$ is not
monotone in~$p$:
\texttt{paley\_alpha\_not\_monotone : $\exists$ p q, p <
      q $\wedge$ $\exists$ a b, independenceNumber (paley p) a
      $\wedge$ independenceNumber (paley q) b $\wedge$ a > b}}}

The proofs rely on cutting-planes (\texttt{pol}) and reverse unit
propagation (\texttt{rup}) steps; the largest instance (2526 edges,
about 63\,000 proof lines) verifies in about 3.5 minutes.
The pigeonhole example in Table~\ref{tab:applications} additionally
exercises the \texttt{red} rule for symmetry breaking.

\begin{table}[tbh]
\centering
\caption{Independent set on Paley graphs. $p$: prime,
$\alpha$: independence number. Vars: PB variables, Cstrs: PB
constraints. Direct: explicit proof-term construction.
Reflect: reflection-based checker. TO = timeout at 60\,000\,$ms$.}
\label{tab:benchmarks}
\av{\medskip}
\cv{\small}
\begin{tabular}{@{}r@{\quad}r@{\quad}r@{\quad}r@{\quad}r@{\quad}r@{\quad}r@{\quad}r@{\quad}r@{}}
\toprule
$p$ & $\alpha$ & Vars & Cstrs & Solve\,($ms$) & Elab\,($ms$) & Lines & Direct\,($ms$) & Reflect\,($ms$) \\
\midrule
13 & 3 & 13 & 40 & 59 & 12 & 35 & 1\,150 & 802 \\
17 & 3 & 17 & 69 & 38 & 4 & 66 & 2\,310 & 802 \\
29 & 4 & 29 & 204 & 53 & 5 & 233 & 19\,922 & 926 \\
37 & 4 & 37 & 334 & 105 & 6 & 554 & TO & 1\,314 \\
41 & 5 & 41 & 411 & 131 & 6 & 704 & TO & 1\,536 \\
53 & 5 & 53 & 690 & 421 & 13 & 1\,848 & TO & 3\,494 \\
61 & 5 & 61 & 916 & 850 & 24 & 4\,151 & TO & 6\,743 \\
73 & 5 & 73 & 1\,315 & 3\,145 & 64 & 9\,808 & TO & 20\,616 \\
89 & 5 & 89 & 1\,959 & 17\,166 & 234 & 36\,795 & TO & 96\,234 \\
97 & 6 & 97 & 2\,329 & 21\,673 & 272 & 43\,214 & TO & 115\,197 \\
101 & 5 & 101 & 2\,526 & 29\,267 & 385 & 62\,924 & TO & 200\,460 \\
\bottomrule
\end{tabular}
\end{table}

\av{A noteworthy byproduct of this evaluation is a
formally verified proof that $\alpha(\mathrm{Paley}(p))$ is not
monotone in~$p$:
$\alpha(\mathrm{Paley}(97)) = 6 > 5 = \alpha(\mathrm{Paley}(101))$.
Here, each equality combines the \veripb upper bound with a lower
bound from an explicit witness verified by \texttt{native\_decide}.}\par\medskip\noindent
\begin{minipage}{\linewidth}
\begin{lstlisting}
theorem paley_alpha_not_monotone :
    $\exists$ p q : Nat, p < q $\wedge$ $\exists$ a b : Nat,
      independenceNumber (paley p) a $\wedge$
      independenceNumber (paley q) b $\wedge$ a > b
\end{lstlisting}
\end{minipage}\medskip

As noted in Section~\ref{sec:formalization}, the explicit proof-term
approach does not scale: the Direct column shows that
$\mathrm{Paley}(29)$ already takes 20\,s, and all instances with
$p \geq 37$ exceed the 60\,s timeout.
In contrast, the reflection-based checker verifies
$\mathrm{Paley}(53)$ in 3.5\,s and $\mathrm{Paley}(101)$ in
200\,s. The bottleneck appears to be the compiled checker execution rather
than Lean's kernel, which only checks the soundness theorem application.

Beyond independence numbers of Paley graphs, we provide trusted
encodings for seven additional combinatorial problems. These
applications demonstrate the generality of the verified encoding
approach. The underlying theorems are well known; the contribution
lies in the end-to-end formal verification within Lean.
Table~\ref{tab:applications} reports the same pipeline metrics as
Table~\ref{tab:benchmarks}.

\begin{table}[tbh]
\centering
\caption{Additional verified combinatorial results, using the
  reflection-based checker. Column
headers as in Table~\ref{tab:benchmarks}.
Dashes indicate a hand-crafted proof (no solver call).}
\label{tab:applications}
\cv{\small}
\av{\medskip}
\begin{tabular}{@{}l@{\quad}r@{\quad}r@{\quad}r@{\quad}r@{\quad}r@{\quad}r@{}}
\toprule
Problem & Vars & Cstrs & Solve\,($ms$) & Elab\,($ms$) & Lines & Reflect\,($ms$) \\
\midrule
Langford $\neg L(2,6)$
  & 72 & 463 & 152 & 12 & 223 & 1\,221 \\
Langford $\neg L(2,9)$
  & 162 & 1\,867 & 12\,000 & 180 & 44\,342 & 155\,000 \\
Schur $S(2)=4$
  & 6 & 12 & 35 & 4 & 17 & 843 \\
Schur $\neg S_3(14)$
  & 42 & 161 & $<$100 & 4 & 209 & 880 \\
VdW $W(2,3)=9$
  & 10 & 32 & 37 & 4 & 19 & 867 \\
VdW $\neg W_4(35)$
  & 36 & 374 & $<$100 & 4 & 429 & 870 \\
Ramsey $R(3,3)=6$
  & 15 & 40 & 37 & 4 & 31 & 894 \\
Ramsey $\neg R(3,4)$ on $K_9$
  & 36 & 210 & 3\,000 & 30 & 25\,194 & 26\,000 \\
Eq.~col.~$\chi_{eq}(K_{3,3,1})\!=\!5$
  & 28 & 117 & 72 & 5 & 490 & 957 \\
PHP $\neg\mathrm{PHP}(3,2)$
  & 6 & 7 & --- & --- & 22 & 815 \\
Bin packing $12/5/14$
  & 60 & 17 & $<$100 & 4 & 687 & 843 \\
\bottomrule
\end{tabular}
\end{table}

\paragraph{Langford pairing.}
A \emph{Langford pairing} of order~$n$ places each integer
$k \in \{1,\ldots,n\}$ into a sequence of length~$2n$ such that the
two occurrences of~$k$ are exactly $k+1$ positions apart. Langford
pairings exist if and only if $n \equiv 0$ or $3 \pmod{4}$.
\begin{lstlisting}
theorem langford6_impossible : $\lnot$hasLangfordPairing 6
theorem langford9_impossible : $\lnot$hasLangfordPairing 9
\end{lstlisting}
The impossibility of $L(2,6)$ is proved from a 223-line kernel proof.
The larger instance $L(2,9)$ (162~variables, 1\,867~constraints)
requires a 44\,342-line proof and takes about 155\,s to verify,
demonstrating that the approach scales to proofs with tens of
thousands of steps.

\paragraph{Schur number.}
A 2-coloring of $\{1,\ldots,n\}$ is \emph{Schur-free} if no
monochromatic triple $(a, b, a{+}b)$ exists. The Schur number $S(2)$
is the largest~$n$ for which there exists such a coloring.
\begin{lstlisting}
theorem schur_number_2 : schurNumber 4
theorem schur14_impossible : $\lnot$hasKSchurFreeColoring 3 14
\end{lstlisting}
The $S(2)$ lower bound uses the witness coloring
$\{1,4\} \mapsto \mathrm{true}$, $\{2,3\} \mapsto \mathrm{false}$.
The upper bound ($\{1,\ldots,5\}$ has no Schur-free coloring) is
verified from a 17-line kernel proof. For
$S(3) = 13$, the encoding generalizes to $k$-colorings using
a one-hot representation: $k$ binary variables per element plus
at-least-one and no-monochromatic-triple constraints per color.
The resulting instance (42~variables, 161~constraints) is verified
from a 209-line proof.

\paragraph{Van der Waerden number.}
A 2-coloring of $\{1,\ldots,n\}$ is \emph{AP-free} if no
monochromatic 3-term arithmetic progression $(a, a{+}d, a{+}2d)$
exists. The Van der Waerden number $W(2,3)$ is the largest~$n$
for which there exists such a coloring.
\begin{lstlisting}
theorem vdw_number_2_3 : vanDerWaerdenNumber 8
theorem vdw35_impossible : $\lnot$hasKAPFreeColoring 4 35
\end{lstlisting}
The $W(2,3)$ lower bound uses the witness $\texttt{RRBBRRBB}$; the
upper bound ($\{1,\ldots,9\}$ has no AP-free coloring) is verified
from a 19-line kernel proof. The generalized encoding supports
$k$-term arithmetic progressions. For $W(2,4) = 35$, the upper bound
($\{1,\ldots,35\}$ has no 2-coloring avoiding monochromatic 4-APs)
is verified from a 429-line proof over 36~variables.

\paragraph{Ramsey number.}
A 2-coloring of the edges of the complete graph~$K_n$ is
\emph{Ramsey-free} if it contains no monochromatic triangle. The
Ramsey number $R(3,3)$ is the smallest~$n$ such that every 2-coloring
of~$K_n$ contains a monochromatic triangle.
\begin{lstlisting}
theorem ramsey_3_3 : ramseyNumber 5
theorem ramsey9_34_impossible :
    $\lnot$hasAsymRamseyFreeColoring 9 3 4
\end{lstlisting}
The $R(3,3)$ lower bound uses the pentagon~($C_5$) coloring of~$K_5$.
The upper bound---that $K_6$ admits no Ramsey-free coloring---is
verified from a 31-line kernel proof. The generalized encoding
supports asymmetric Ramsey numbers $R(s,t)$: forbid red~$K_s$ and
blue~$K_t$ separately. For $R(3,4) \leq 9$, the instance has
36~edge variables and 210~constraints (84~triangle and
126~$K_4$ constraints); the 25\,194-line proof verifies in
about 26\,s.

\paragraph{Equitable chromatic number.}
A proper $k$-coloring of a graph~$G$ is \emph{equitable} if all color
classes have size in $\{\lfloor n/k \rfloor, \lceil n/k \rceil\}$.
The equitable chromatic number $\chi_{eq}(G)$ is the smallest
such~$k$. The balance constraints are natively pseudo-Boolean: they
express cardinality bounds as linear inequalities, which would require
auxiliary encoding variables in pure SAT.
\begin{lstlisting}
theorem k331_eq_chromatic : equitableChromaticNumber k331 5
\end{lstlisting}
The complete tripartite graph $K_{3,3,1}$ (7~vertices, 15~edges) has
standard chromatic number $\chi = 3$ but equitable chromatic number
$\chi_{eq} = 5$, demonstrating a gap of~2. The upper bound uses an
explicit 5-coloring witness; the lower bound
($\neg\mathit{hasEquitableColoring}\ K_{3,3,1}\ 4$) is verified via
the reflection checker from a 490-line kernel proof with 96
proof-by-contradiction subproofs.

\paragraph{Pigeonhole principle.}
The \emph{pigeonhole principle} $\mathrm{PHP}(m,n)$ states that $m$~pigeons
cannot be placed into $n$~holes with at most one pigeon per hole when
$m > n$. The proof of $\neg\mathrm{PHP}(3,2)$ uses the \texttt{red}
rule to add a symmetry-breaking constraint (without loss of generality,
pigeon~1 goes to hole~1) via a cyclic substitution witness.
This is the only example exercising the redundance rule.
\begin{lstlisting}
theorem php32_red_reflect : formulaUnsat
\end{lstlisting}

\paragraph{Bin packing.}
Given $n$ items with sizes $s_0, \ldots, s_{n-1}$ and $m$ bins of
capacity~$C$, the \emph{bin packing} problem asks whether all items
can be assigned to bins without exceeding capacity. The capacity
constraint $\sum_i s_i \cdot x_{i,j} \leq C$ is a single PB
inequality; in CNF it requires a sequential weight counter
encoding~\cite{PBLib} with $O(n \cdot C)$ auxiliary variables per bin.
Our instance has 12~items (sizes $[10,9,8,8,6,5,4,4,4,4,4,4]$),
5~bins, and $C = 14$. The total size equals $5 \times 14 = 70$,
so total weight alone does not refute feasibility. The instance is
not AMO\nobreakdash-reducible: pairs like $10 + 4 = 14$ and triples
like $4 + 4 + 4 = 12$ fit in a single bin. Unsatisfiability follows
from a rounding argument: dividing each capacity constraint by~4
shows that each bin holds at most 3~``slots,'' heavy items ($\geq 8$)
consume~2 slots and light items ($\geq 4$) consume~1 slot, giving a
total demand of $2 \times 4 + 8 = 16 > 15 = 3 \times 5$. The PB encoding consists of just 17~constraints over 60~variables;
the VeriPB kernel proof has 687~lines. A CNF encoding of the same
instance via the standard sequential weight counter~\cite{PBLib}
requires 960~variables (900~auxiliary) and 2\,617~clauses, and
this encoding overhead grows linearly with the capacity~$C$.
\begin{lstlisting}
theorem bp12_5_impossible : $\lnot$hasPacking inst12_5
\end{lstlisting}

\section{Related Work}
\label{sec:related}
\SPACE
\cakepb~\cite{GochtMMNOT24} provides end-to-end verified subgraph
solving in HOL4/CakeML, including both a PB proof checker and verified
graph-problem encoders compiled to a standalone binary. \av{As discussed in
Section~\ref{sec:intro}, the approaches are complementary:} \cakepb
targets standalone verification with a small trusted computing base,
while \pblean yields Lean theorems that can serve as lemmas in
larger formal developments. Running \cakepb on our Paley instances,
it checks $\mathrm{Paley}(101)$ in under 1\,s; \pblean takes
about 200\,s. This roughly $200\times$ overhead is the cost of
producing a Lean theorem rather than a standalone verdict.
Several extensions of the \veripb framework are known, certifying
MaxSAT preprocessing~\cite{IhalainenOTBJMN24}, solution-improving
search~\cite{BergBNOPV24}, dynamic programming and decision
diagrams~\cite{DemirovicMMNOS24}, and pseudo-Boolean
optimization~\cite{KoopsBMNOTV25}.

An alternative to native PB proof checking is to compile PB reasoning
to clausal proofs. Bryant et al.~\cite{BryantBH22} showed that this is
possible via BDD conversion, and their approach efficiently handles pigeonhole and
parity instances via extension variables. Direct \veripb
import provides a simpler pipeline that works with the native proof
format of PB solvers.

Approaches for SAT certificate checking in theorem provers include
Heule et al.'s verified \lrat checker in ACL2~\cite{HeuleHKW17},
Lammich's GRAT in Isabelle/HOL~\cite{Lammich17}, and
Lean's \texttt{FromLRAT}~\cite{FromLRAT}, whose architecture our
integration closely follows. Heule and
Scheucher~\cite{HeuleS24} established the empty hexagon theorem via SAT
solving with verified proof checking; Subercaseaux
et~al.~\cite{SubercaseauxNGCCH24} formalized the encoding in Lean and
verified the reduction from geometry to SAT.

Rather than checking proof certificates, one can verify the solver
itself. Blanchette et al.~\cite{BlanchetteFLW18} formalized CDCL in
Isabelle/HOL as a chain of refinements from abstract calculus to
executable code; Fleury and Lammich~\cite{FleuryL23} extended this to
IsaSAT, a verified SAT solver.
\av{SMTCoq~\cite{EkiciMTKKRB17} imports SMT witnesses into Coq. Besson's
\texttt{micromega}~\cite{Besson06} uses reflexive certificate checking
for linear arithmetic in Coq, based on Farkas' Lemma rather than cutting
planes, and does not import external PB proofs. A related
project~\cite{LeanCSP} formalizes general CSP definitions in Lean~4
with verified translations to MiniZinc and SMT-LIB solver formats.
\pblean extends verified PB checking to Lean~4, complementing existing
SAT checking in Isabelle and PB checking in HOL4.}\cv{SMTCoq~\cite{EkiciMTKKRB17} imports SMT witnesses into Coq;
Besson's \texttt{micromega}~\cite{Besson06} uses reflexive certificate
checking for linear arithmetic in Coq but does not import external PB
proofs. A related project~\cite{LeanCSP} formalizes general CSP
definitions in Lean~4 with verified translations to solver formats.
\pblean extends verified PB checking to Lean~4, complementing existing
SAT checking in Isabelle and PB checking in HOL4.}

\section{Conclusion} \label{sec:conclusion}
\SPACE
We propose a Lean~4 integration for \veripb
pseudo-Boolean proof certificates.
The key to \pblean's scalability is reflection: a Boolean checker
with a fully proved soundness theorem, evaluated natively by the kernel.
As demonstrated in Section~\ref{sec:evaluation}, this verifies
proofs with tens of thousands of steps that would exhaust memory under
explicit proof construction. Our largest instance, $\mathrm{Paley}(101)$
with about 63\,000 proof lines, verifies in about 3.5 minutes.

With trusted encodings, users obtain end-to-end verified combinatorial
results: the independence number of $\mathrm{Paley}(101)$ is a Lean
proposition, not an external verdict. This
composability, not directly available with standalone verified
checkers, enables PB reasoning to serve as a building block in larger
formalizations.

Future work includes a \texttt{veripb} tactic for proof automation and
integration with other Lean formalizations.

\av{
\section*{Acknowledgments}
Research was carried out within the
Cluster of Excellence \emph{Bilateral AI} (10.55776/COE12) of the
Austrian Science Funds (FWF).}
\bibliographystyle{plainurl}
\bibliography{references}

\end{document}